# Observation of insulator-metal transition in EuNiO$_3$ under high pressure

R Lengsdorf[1], A Barla[2], J A Alonso[3], M J Martinez-Lope[3], H Micklitz[1] and M M Abd-Elmeguid[1]

[1]II. Physikalisches Institut, Universität zu Köln, Zülpicher Str. 77, 50937 Köln, Germany
[2]European Synchrotron Radiation Facility (ESRF), 6 Rue Jules Horowitz, BP 220, 38043 Grenoble Cedex 9, France
[3]Instituto de Ciencia de Materiales de Madrid, CSIC, Cantoblanco, E-28049 Madrid, Spain

**Abstract**

The charge transfer antiferromagnetic ($T_N$ = 220 K) insulator EuNiO$_3$ undergoes at ambient pressure a temperature-induced metal-insulator (MI) transition at $T_{MI}$ = 463 K. We have investigated the effect of pressure (up to $p \approx$ 20 GPa) on the electronic, magnetic and structural properties of EuNiO$_3$ using electrical resistance measurements, $^{151}$Eu nuclear resonance scattering of synchrotron radiation and x-ray diffraction, respectively. With increasing pressure we find at $p_c$ = 5.8 GPa a transition from the insulating state to a metallic state, while the orthorhombic structure remains unchanged up to 20 GPa. The results are explained in terms of a gradual increase of the electronic bandwidth with increasing pressure which results in a closing of the charge transfer gap. It is further shown that the pressure-induced metallic state exhibits magnetic order with a lower value of $T_N$ ($T_N \approx$ 120 K at 9.4 GPa) which disappears between 9.4 GPa and and 14.4 GPa.

## 1. Introduction

In recent years, there has been considerable interest in investigating the unusual magnetic and electronic properties of rare earth (R) nickelates $RNiO_3$ perovskites, in particular the metal-insulator (MI) transition [1]. This is due to the fact that unlike the manganates, $RNiO_3$ oxides, are placed at the boundary separating low-$\Delta$-metals ($\Delta$ is the charge transfer energy) and charge transfer insulators. An exciting consequence of this is that the occurrence of the metal-insulator transition in $RNiO_3$ series requires neither electron nor hole doping. The MI transition can be induced as a function of temperature and the transition temperature ($T_{MI}$) increases with decreasing the size of the $R^{3+}$ ions. For R = La the system remains metallic down to low temperatures, whereas for R = Pr, Nd, Sm and Eu electron localization occurs at $T_{MI}$ = 130 K, 200 K, 400 K and 463 K, respectively [2-4]. Above $T_{MI}$, the crystal structure of all $RNiO_3$ (R ≠ La) perovskites is orthorhombic (space group Pbnm) and no change of the lattice symmetry has been observed in these compounds on going from the metallic to the insulating state: only a slight expansion of the unit cell volume takes place across the MI transition as a result of a small increase of the Ni-O bond length ($d_{Ni-O}$) and a simultaneous decrease of the Ni-O-Ni bond angle ($\theta$) [4,5]. In fact, the value of $T_{MI}$ has been found to be related to the steric effect induced by the size of the $R^{3+}$ ion via the change of $d_{Ni-O}$ and $\theta$ which determines the degree of overlapping of Ni (3d) and O (2p) orbitals and hence the electronic bandwidth.

In this respect, external pressure can be very efficient, e.g. by modifying the effective bandwidth ($W$) of Ni (3d) states through changing the Ni-O bond length and/or the Ni-O-Ni bond angle, thereby providing a unique tool to tune electronic and magnetic properties by "bandwidth control" of the system. However, until now, little is known about the effect of pressure on the electronic, magnetic and structural properties of the $RNiO_3$ series. Only the pressure dependence of the $T_{MI}$ of $PrNiO_3$ and $NdNiO_3$ has been investigated using electrical resistance measurements up to 1.6 GPa [6, 7]. These studies showed a strong reduction of $T_{MI}$ with increasing pressure, i.e. a stabilization of the metallic state which has been explained as a result of an increase of $W$ with increasing pressure due to a corresponding decrease of $d_{Ni-O}$ and a simultaneous increase of $\theta$ [7]. This explanation has been experimentally supported by high pressure neutron diffraction experiments on $PrNiO_3$ [8].

Another interesting aspect of RNiO$_3$ compounds is that for larger R$^{3+}$ ions (R = Pr, Nd) the MI transition is accompagnied by an antiferromagnetic (AF) ordering of the Ni sublattice (i.e. $T_N \approx T_{MI}$). However, for smaller R$^{3+}$ ions (R = Sm — Lu) $T_N$ is much smaller than $T_{MI}$ (e.g. for EuNiO$_3$ $T_N$ = 220 K and $T_{MI}$ = 463 K) suggesting that the MI transition is not connected with the onset of AF ordering. On the other hand, the observation of very large $^{16}$O – $^{18}$O isotope shifts of $T_{MI}$ in RNiO$_3$ (R = Pr, Nd, Sm and Eu) demonstrates the importance of electron-lattice coupling in these systems [9].

Towards a better understanding of the mechanism of the MI transition in RNiO$_3$ compounds we have investigated the effect of pressure on the electronic, magnetic and structural properties of EuNiO$_3$. This is the first high pressure study of a member of RNiO$_3$ series in which $T_{MI}$ and $T_N$ are well apart ($T_N << T_{MI}$). Using electrical resistance measurements, $^{151}$Eu Nuclear Forward Scattering (NFS) of synchrotron radiation and x-ray diffraction we find a pressure-induced insulator to metal transition at a critical pressure $p_c$ = 5.8 GPa, while the orthorhombic structure remains unchanged. We further show that the metallic state exhibits magnetic order with a reduced value of $T_N$ ($T_N \approx$ 120 K at 9.4 GPa).

## 2. Experimental details

A single phase polycrystalline sample of EuNiO$_3$ was prepared under an oxygen pressure of 200 bar. Details of the preparation and characterization are published elsewhere [4, 10]. The pressure dependence of the lattice constants at 300 K up to about 20 GPa was measured on powdered samples by energy dispersive x-ray diffraction (EDX) at HASYLAB using the diamond anvil cell (DAC) technique. The same type of DAC has been used for conventional four-terminal electrical resistance measurements up to about 20 GPa between 4.2 K and 300 K. The $^{151}$Eu high pressure nuclear forward scattering (NFS) measurements were performed at the undulator beamline ID22N of the European Synchrotron Radiation Facility (ESRF) in Grenoble [11]. The setup for high pressure NFS experiments is described elsewhere [12].

## 3. Results and discussion

*3.1 pressure-induced insulator-metal transition*

Figure 1 shows the temperature dependence of the electrical resistance $R(p,T)$ between 4.2 K and 300 K at different pressures up to 22 GPa. Since the value of $T_{MI}$ for EuNiO$_3$ at ambient pressure is much higher than room temperature ($T_{MI}$ = 463 K [4]), one observes an insulating type behaviour (i.e. $dR/dT < 0$) at $p = 0$ GPa (100 K $\leq$ T $\leq$ 300 K). With increasing pressure 0 $\leq p \leq$ 5.8 GPa, we observe a strong reduction of $R(T)$ and a gradual decrease of the magnitude of $dR/dT$. The small anomaly observed around 50 K is found to be almost pressure independent (s. figure 1). The origin of such an anomaly is not known at the moment and will not be further discussed. At $p \geq$ 5.8 GPa we find a metallic like behavior ($dR/dT > 0$) in the whole temperature range and further decrease of $R(T)$ with increasing pressure. For example, at $T$ = 125 K the resistance decreases by more than 4 orders of magnitude between 0 GPa and 5.8 GPa. This clearly indicates a continuous pressure-induced insulator-metal transition in EuNiO$_3$ at $p_c$ = 5.8 GPa. In connection to this we would like to mention that we obtained in the metallic state ($p \geq$ 5.8 GPa) an approximate value of the specific resistance ($\rho$) of a few hundreds µΩcm at 4.2 K. This value is higher than that typical for metallic systems, but is comparable with those reported for other members of the RNiO$_3$ series. For example, the value of $\rho$ in the metallic state of PrNiO$_3$ and NdNiO$_3$ amounts to $\rho \approx$ 0.4 mΩcm and 1.3 mΩcm at 300 K, respectively [13,14].

In order to investigate whether the pressure-induced MI transition in EuNiO$_3$ is connected with a structural phase transition or not, we now consider the pressure dependence of the lattice parameters (*a, b* and *c*) and the volume of the orthorhombic unit cell as obtained from our EDX diffraction at 300 K. As shown in Figure 2, we find within the accuracy of the measurements no <u>discontinuity</u> in the pressure dependence of the lattice parameters *a*, *b* and *c* (and the volume), thus showing no indication of a structural phase transition. This figure further reveals that the pressure variation of the lattice parameters *a*, *b* and *c* is quite different: with increasing pressure *b* decreases more rapidly than *a* and *c*, and the value of *c* nearly saturates around 20 GPa. Such a change of *a*, *b* and *c* indicates a gradual decrease of the orthorhombic distortion of the unit cell with increasing pressure. This results in a decrease / increase of the Ni-O-Ni tilt / bond angle of the $NiO_6$ octahedron. The saturation behavior of the lattice parameter *c* around 20 GPa can be taken as an indication that the orthorhombic structure is approaching a structural phase transition with higher symmetry (e.g. orthorhombic to rhombohedral transition). Indeed, preliminary x-ray diffraction measurements at pressure up to ~ 45 GPa indicate the occurrence of such a phase transition above 25 GPa [15]. Thus, the above mentioned high pressure results clearly show that the transition from the insulating state to a metallic state at $p_c$ = 5.8 GPa is not connected with a change of the symmetry of the orthorhombic structure. We attribute the pressure-induced MI transition in $EuNiO_3$ to a gradual decrease of the Ni-O bond length and a simultaneous increase of the Ni-O-Ni bond angle with increasing pressure which cause an increase of the effective 3d bandwidth and thereby closing the charge transfer gap. This interpretation is similar to that reported in the case of $PrNiO_3$ and $NdNiO_3$ to explain the strong decrease of $T_{MI}$ with increasing pressure [7] which has been experimentally confirmed by high-pressure neutron diffraction measurements on $PrNiO_3$ [8].

In this context, it is to be mentioned that our *R(T, p)* data shown in Figure 1 does not show any significant anomaly at $T_{MI}$ between ambient pressure and $p_c$ = 5.8 GPa, as in the case of $PrNiO_3$ and $NdNiO_3$ [6,7]. This can be explained by the fact that in $EuNiO_3$ the anomaly in the resistivity at $T_{MI}$ (463 K at ambient pressure) is considerably weaker than that observed in $PrNiO_3$ and $NdNiO_3$. In addition, since the value of $T_{MI}$ in $EuNiO_3$ is much higher than the temperature range of our measurements (4.2 K ≤ T ≤ 300 K), the initial change of $T_{MI}$ with pressure for $p < p_c$ (in the insulating state) is not experimentally accessible. Nevertheless, our data clearly shows a metallic state at 5.8 GPa up to 300 K, i.e. a temperature of T = 300 K can be considered as a lower limit of the value of $T_{MI}$. Assuming a linear decrease of $T_{MI}$ with increasing pressure between ambient pressure and 5.8 GPa, we obtain a crude estimation of the value of $\partial T_{MI}/\partial p \approx -30$ KGPa$^{-1}$ which is comparable with those reported for PrNiO3 and NdNiO3 ($\partial T_{MI}/\partial p = -42$ KGPa$^{-1}$) [6]. This is consistent with our conclusion that the mechanism involved in the pressure-induced insulator-metal transition in EuNiO3 is essentially similar to that suggested for PrNiO3 and NdNiO3 [7].

*3.2 magnetic ordering in the metallic state*

The last point to be discussed is the effect of pressure on the magnetic state of $EuNiO_3$ and its possible relationship to the observed pressure-induced MI transition. This has been achieved by performing $^{151}$Eu nuclear forward scattering (NFS) of synchrotron radiation on $EuNiO_3$ under high pressure. This technique allows one to probe the magnetic state of the Ni-sublattice under high pressure via the induced magnetic hyperfine (*hf*) field $B_{ind}$ at the $^{151}$Eu nuclei which results from the ordered Ni (3d) moments. $B_{ind}$ originates from the Ni-Eu exchange field due to the admixture of the excited $Eu^{3+}$ states (e.g. $^7F_1$) into the nonmagnetic ground state ($^7F_0$) [16]. Some selected $^{151}$Eu NFS spectra of $EuNiO_3$ collected at $T = 3$ K for different pressures (0 GPa, 9.4 GPa and 14.4 GPa) are shown in figure 3a. The temperature dependence of NFS spectra measured at 9.4 GPa is displayed in figure 3b. The fits to these spectra were performed using the program package CONUSS [17] by incorporating the full dynamical theory of nuclear resonance scattering, including the diagonalization of the complete hyperfine Hamiltonian. Starting with the results at ambient pressure and at $T = 3$ K (s. figure 3 a), i.e. in the insulating state, the spectrum can be best fitted assuming an induced *hf* field $B_{ind} = 1.4(2)$ T combined with a small quadrupole interaction ($\Delta E_Q = eQ_gV_{zz} = 6.0(4)$ mm/s) which is due to the orthorhombic point symmetry of the Eu lattice sites. The induced *hf* field is due to the Eu-Ni exchange field of the 4 parallel spins surrounding the Eu atoms as reported by neutron diffraction data on $EuNiO_3$ [18]. At 9.4 GPa and $T = 3$ K, i.e. in the metallic state, we find within the accuracy of measurements no significant change of the values of $B_{ind} = 1.2(5)$ T and $eQV_{zz} = 6.2(14)$ mm/s. The difference in the shape of the NFS spectra at 9.4 GPa and 0 GPa is due to the different thicknesses of the sample measured at ambient pressure and in the DAC. This clearly indicates that the metallic state at $p = 9.4$ GPa is magnetically ordered at $T = 3$ K. At 14.4 GPa ($T = 3$ K) no induced magnetic *hf* field could be detected and only a quadrupole splitting is left ($eQ_gV_{zz} = 6.6(4)$ mm/s), suggesting the disappearance of magnetic ordering between 9.4 GPa and 14.4 GPa. In order to have an estimate of the transition temperature of the magnetically ordered state at 9.4 GPa we have measured NFS spectra at different temperatures between 300 K and 3 K (s. figure 3b. The NFS spectra at 300 K and 150 K could be fitted by assuming a pure quadrupole splitting ($\Delta E_Q = 6.2(4)$ mm/s), indicating a paramagnetic state. In contrast, for the analysis of the spectrum at 50 K we had to assume a combined quadrupole and magnetic *hf* interaction. We obtain values for $\Delta E_Q$ = 6.4(4) and $B_{ind}$ = 1.3(3) T, which are similar to those found at 3 K (s. above). This indicates that at 50 K the magnetically ordered state is nearly saturated. The analysis of the temperature dependence of the spectra at 9.4 GPa shows that the transition temperature is around 120 K, which is considerably lower than that at ambient pressure ($T_N$ = 220 K). Unfortunately, the $^{151}$Eu NFS technique provides no information about the type of magnetic structure. Therefore, we can not drive any conclusion about a possible change of the magnetic structure which may be associated with a change of the orbital state. In any case, it would be interesting to determine the pressure-induced change of $T_N$ in the insulating state and across the metal-insulator transition. In particular, it is highly desired to determine the critical pressure at which the magnetic ordering vanishes ($T_N \rightarrow 0$), i.e. the magnetic quantum critical point of EuNiO$_3$. This requires systematic and accurate $^{151}$Eu high pressure NFS experiments on EuNiO$_3$. Such experiments are underway.

## 4. Conclusion

We have investigated the effect of pressure on the electronic, magnetic and structural properties of the charge transfer antiferromagnetic ($T_N$ = 220 K) insulator EuNiO$_3$ which undergoes at ambient pressure a temperature-induced metal insulator transition at $T_{MI}$ = 463 K. We find a pressure-induced insulator metal transition at $p \geq 5.8$ GPa, while the lattice structure does not change up to about 20 GPa. The metallic state is found to exhibit magnetic ordering with a lower value of $T_N$ ($T_N \approx 120$ K at 9.4 GPa) which disappears between 9.4 GPa and 14.4 GPa.

**Acknowledgements**

The authors would like to thank O Leupold and H C Wille for assistance during the measurements at the ESRF. M M A would like to thank D I Khomskii, H L Tjeng and J P Sanchez for useful discussions. This work was supported by the Deutsche Forschungsgemeinschaft through SFB 608.

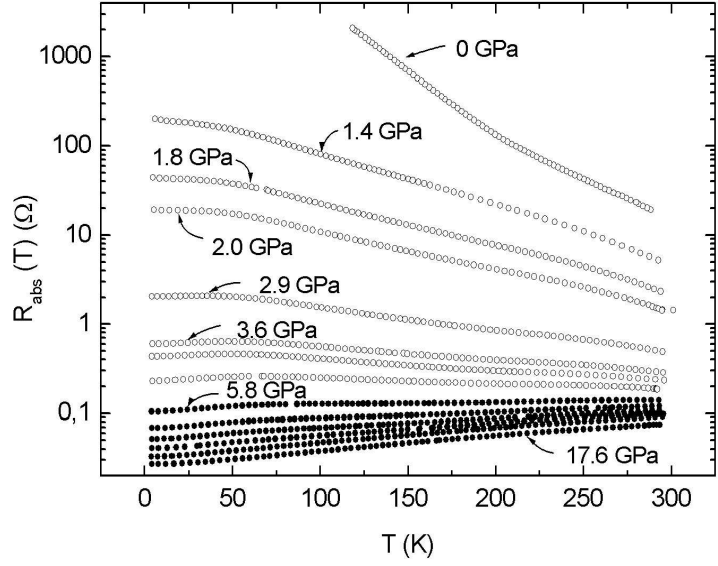

**Figure 1.** Temperature dependence of the electrical resistance $R(p,T)$ of EuNiO$_3$ at different pressures up to 17.6 GPa. Pressure values between 5.8 GPa and 17.6 GPa are 7.2, 8.5, 10.1 and 13.2 GPa, respectively.

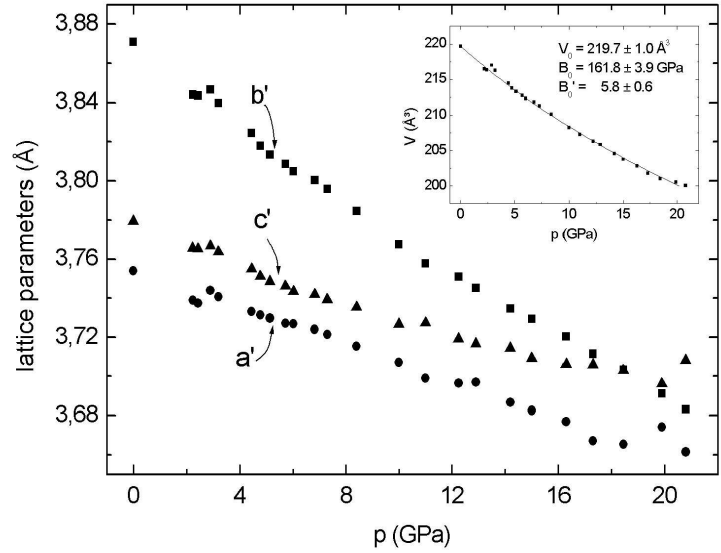

**Figure 2.** Pressure dependence of the lattice parameters *a*, *b*, and *c* ($a = a'\sqrt{2}$, $b = b'\sqrt{2}$, and $c = 2\,c'$) of EuNiO$_3$ as obtained from high pressure x-ray diffraction measurements at 300 K. The inset shows the variation of the unit cell volume with pressure. The value of the bulk modulus ($B_0$) and its derivative ($B_0'$) as obtained from the fit of the equation of state using the Birch-Murnaghan equation are $B_0 = 162(4)$ GPa and $B_0' = 6(1)$.

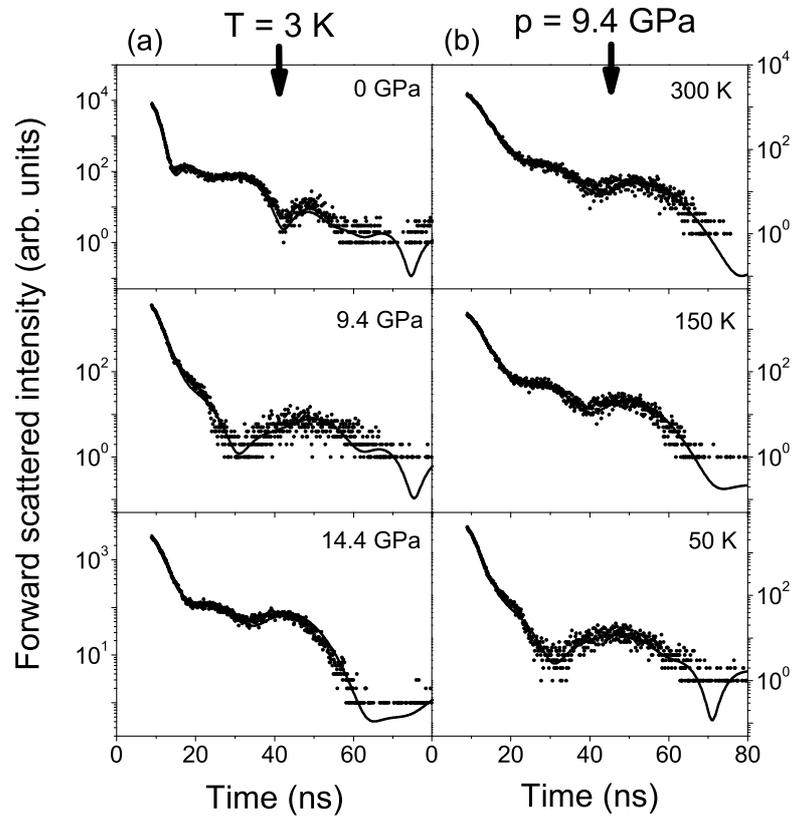

**Figure 3:** $^{151}$Eu NFS spectra of EuNiO$_3$ (a) measured at $T = 3$ K and at different pressures of 0 GPa, 9.4 GPa and 14.4 GPa; and (b) measured at 9.4 GPa and at different temperatures of 300 K, 150 K and 50 K. The lines through the data points are fits (s. text).